\documentclass{iopjournal}

\usepackage{amsmath}
\usepackage{amssymb}
\usepackage{amsfonts}
\usepackage{mathtools}

\usepackage{amsthm}

\usepackage{cleveref}

\usepackage[numbers]{natbib}

\theoremstyle{plain}
\newtheorem{theorem}{Theorem}[section]
\newtheorem{proposition}[theorem]{Proposition}

\theoremstyle{definition}
\newtheorem{definition}[theorem]{Definition}

\theoremstyle{remark}

\newcommand{\Hobj}{\mathcal{H}_{\mathrm{obj}}}
\newcommand{\Hdata}{\mathcal{H}_{\mathrm{data}}}

\crefname{theorem}{Theorem}{Theorems}
\crefname{proposition}{Proposition}{Propositions}
\crefname{lemma}{Lemma}{Lemmas}
\crefname{corollary}{Corollary}{Corollaries}
\crefname{definition}{Definition}{Definitions}
\crefname{assumption}{Assumption}{Assumptions}
\crefname{remark}{Remark}{Remarks}

\begin{document}

\articletype{Paper} 

\title{Information Geometry of Imaging Operators}

\author{Charles Wood$^1$\orcid{0009-0007-4614-2695}}

\affil{$^1$Future Technology Centre, School of Electrical and Mechanical Engineering, University of Portsmouth, PO1 3HE, UK}

\email{charles.wood@port.ac.uk}


\begin{abstract}

\medskip
Imaging systems are represented as linear operators, and their singular value spectra describe the structure recoverable at the operator level. Building on an operator-based information-theoretic framework, this paper introduces a minimal geometric structure induced by the normalised singular spectra of imaging operators. By identifying spectral equivalence classes with points on a probability simplex, and equipping this space with the Fisher--Rao information metric, a well-defined Riemannian geometry can be obtained that is invariant under unitary transformations and global rescaling. The resulting geometry admits closed-form expressions for distances and geodesics, and has constant positive curvature. Under explicit restrictions, composition enforces boundary faces through rank constraints and, in an aligned model with stated idealisations, induces a non-linear re-weighting of spectral states. Fisher–Rao distances are preserved only in the spectrally uniform case. The construction is abstract and operator-level, introducing no optimisation principles, stochastic models, or modality-specific assumptions. It is intended to provide a fixed geometric background for subsequent analysis of information flow and constraints in imaging pipelines.
\end{abstract}

\section{Introduction}
\label{sec:introduction}

\medskip
\noindent
This paper introduces a minimal geometric structure induced by imaging operators when represented through their normalised singular spectra. Singular value spectra are used here as a basis- and scale-invariant summary of linear mode weighting under a fixed discretisation. Throughout this paper, the term ``recoverable'' is used only in this restricted operator-level sense: singular values quantify how strongly an operator amplifies or suppresses its singular modes. No task-dependent, noise-dependent, estimator-dependent, or performance-theoretic notion of recoverability is implied.

\medskip
\noindent\textbf{Standing assumptions.}
An imaging system is modelled as a bounded linear operator between finite-dimensional Hilbert spaces induced by a fixed discretisation.
The squared singular values of this operator are used as a basis- and scale-invariant description of how its action distributes energy across modes.
No task model, noise model, or reconstruction model is assumed, and the term ``recoverable'' is used solely in this operator-level sense of spectral allocation, not in a decision-, estimation-, or performance-theoretic sense.

\medskip
\noindent
The aim of the present work is deliberately limited. We do not construct a geometry on the full space of linear operators, nor do we introduce optimisation principles, stochastic models, algorithmic considerations, or modality-specific assumptions. Instead, we identify a reduced state space associated with spectral equivalence classes and equip it with a canonical information geometry. This geometry is then analysed abstractly, including its behaviour under operator composition.

\medskip
\noindent
The purpose here is to fix one structural element: a metric geometry on spectral operator states. Subsequent developments will build on this fixed structure rather than modify it.

\medskip
\noindent
This work is mathematically self-contained, and it sits within a wider effort to develop information-theoretic descriptions of imaging operators.

\subsection{Abstract interpretation of the induced geometry}
\label{subsec:abstract_interpretation}

\medskip
\noindent
The construction above equips equivalence classes of imaging operators with a precise geometric structure derived solely from their normalised singular spectra. At this level of abstraction, the geometry has a direct interpretation derived from the formal results, without reference to any imaging modality or experiment.

\medskip
\noindent
Each operator is represented by a point $\lambda\in\Delta^{N-1}$ encoding how its total squared singular value is distributed across modes. The Fisher--Rao geometry on $\Delta^{N-1}_{>0}$ therefore quantifies differences between operators in terms of \emph{relative spectral allocation}, not absolute scale or basis-dependent structure. Two operators are geometrically close if, and only if, their singular energy is distributed similarly across modes, irrespective of how those modes are realised.

\medskip
\noindent
Geodesics in this geometry represent the smoothest possible redistribution of spectral weight between two operator classes. The explicit form of the geodesics shows that spectral mass is transported continuously through square-root coordinates, ensuring that no mode is extinguished or created along an interior path. This formalises how the geometry represents continuous changes in spectral allocation, as opposed to discrete rank loss.

\medskip
\noindent
The Fisher–Rao manifold has constant positive curvature, so the space of spectral states is intrinsically non-Euclidean. Linear interpolation is only distinguished in trivial cases; typical interpolants curve through the simplex. This curvature expresses a fundamental incompatibility between simultaneous preservation of all spectral proportions and linear superposition, a property that is independent of any specific operator realisation.

\medskip
\noindent
Operator composition acts on this geometry by inducing transport maps on $\Delta^{N-1}$. The formal results establish that only spectrally uniform stages act as isometries; all nontrivial stages distort distances and, in general, the curvature experienced by spectral states. Rank-deficient stages induce a further geometric effect by forcing states onto boundary faces of the simplex, which are metrically singular. These effects together imply that composition generically reshapes the geometric relationships between operator classes rather than merely translating them.

\medskip
\noindent
Finally, the impossibility results clarify the scope of the representation. Since $\lambda$ forgets singular-vector information, the geometry is deliberately insensitive to operator alignment, commutativity, and spatial structure. The resulting geometry is, therefore, not a geometry of operators themselves, but a geometry of \emph{informational equivalence classes} determined by spectral content alone. 

\medskip
\noindent All interpretive statements made above are consequences of this reduction and of the intrinsic properties of the Fisher--Rao metric on the simplex. This abstract interpretation serves only to clarify what the formal geometry encodes and what it necessarily excludes. No additional assumptions or modelling choices are introduced at this stage.

\subsection{Scope and limitations of the present construction}
\label{subsec:scope_limits}

\medskip
\noindent
The geometric framework developed in this paper is intentionally minimal. Its purpose is to establish, in a mathematically controlled manner, that imaging operators induce a non-trivial information geometry when represented through their normalised singular spectra. To avoid overextension of the formalism, several questions and directions are explicitly excluded from the scope of this work:

\medskip
\noindent
First, this paper does not attempt to define a geometry on the full space of linear operators. As a consequence, no claims are made regarding operator commutativity, spatial localisation of modes, or the geometry of singular subspaces. Any extension to geometries that retain directional or vectorial information is deferred to later work.

\medskip
\noindent
Second, the construction is conditional on a fixed discretisation and a fixed spectral dimension $N$.
No claim of invariance under changes of resolution, sampling, or discretisation is made in this paper.

\medskip
\noindent
Finally, the paper does not attempt to exhaust the geometric consequences of the construction. In particular, we do not study higher-order geometric invariants, global topological properties beyond the simplex structure, or extensions to infinite-dimensional limits. These directions are natural continuations of the present framework.

\medskip
\noindent
In summary, this paper establishes a rigorously defined, invariant information geometry on spectral equivalence classes of imaging operators and analyses its basic behaviour under composition. Extensions that incorporate additional operator structure, stochasticity, optimisation, or physical interpretation are intentionally deferred. 

\section{Operator Spectral States}
\label{sec:spectral_states}

\medskip
\noindent
Throughout this paper we work in the finite-dimensional setting induced by discretisation. All Hilbert spaces are finite-dimensional, and all operators are bounded.

\medskip
\noindent\textbf{Fixed spectral dimension.}
For the remainder of this paper we assume a fixed discretisation and a fixed spectral dimension $N$.
All operators compared in $\mathcal{S}=\Delta^{N-1}$ are therefore represented with the same
\[
N=\min\{\dim\Hobj,\dim\Hdata\}
\]
(after discretisation). When multiple stages are considered (for example, in Section~\ref{sec:composition}), we restrict attention to families of operators whose spectral states lie in this common $N$-simplex.
This ensures that all operators share a single state space and a single information geometry. Dimension-changing refinements are outside the scope of the present paper, and no invariance with respect to changes of discretisation or $N$ is claimed.

\subsection{Imaging operators and singular spectra}

\medskip
\noindent
Let $\mathcal{H}_{\mathrm{obj}}$ and $\mathcal{H}_{\mathrm{data}}$ be finite-dimensional Hilbert spaces representing object and data domains, respectively. An imaging system is represented by a linear operator
\[
O : \mathcal{H}_{\mathrm{obj}} \to \mathcal{H}_{\mathrm{data}}.
\]
We denote by $\{\sigma_i(O)\}_{i=1}^N$ the singular values of $O$, ordered such that
\[
\sigma_1(O) \ge \sigma_2(O) \ge \cdots \ge \sigma_N(O) \ge 0,
\]
where $N=\min\{\dim\mathcal{H}_{\mathrm{obj}},\dim\mathcal{H}_{\mathrm{data}}\}$.

\medskip
\noindent
The singular spectrum is taken as the operator-level carrier of recoverable structure. Here “recoverable” is meant only in this spectral sense. No assumptions are made about the physical origin of $O$ beyond linearity and boundedness. 

\subsection{Definition of the spectral state}

\medskip
\noindent
\begin{definition}[Operator spectral state]
Let $O$ be a non-zero imaging operator with singular values $\{\sigma_i(O)\}_{i=1}^N$. The \emph{spectral state} of $O$ is the vector
\[
\lambda(O) = (\lambda_1,\ldots,\lambda_N),
\qquad
\lambda_i
=
\frac{\sigma_i(O)^2}{\sum_{j=1}^N \sigma_j(O)^2}.
\]
\end{definition}

\noindent By construction, $\lambda_i\ge 0$ and $\sum_i \lambda_i=1$.

\subsection{Invariance properties}

\medskip
\noindent
\begin{proposition}[Unitary invariance]
\label{prop:unitary_invariance}
Let $U$ and $V$ be unitary operators on $\mathcal{H}_{\mathrm{data}}$ and $\mathcal{H}_{\mathrm{obj}}$, respectively. Then
\[
\lambda(UOV)=\lambda(O).
\]
\end{proposition}

\begin{proposition}[Scale invariance]
\label{prop:scale_invariance}
For any scalar $c\in\mathbb{R}\setminus\{0\}$,
\[
\lambda(cO)=\lambda(O).
\]
\end{proposition}

\noindent These invariances imply that $\lambda(O)$ depends only on the equivalence class of $O$ under unitary transformations and isotropic rescaling.

\section{The Spectral Simplex as State Space}
\label{sec:spectral_simplex}

\noindent

\subsection{Definition of the state space}
\medskip
\noindent
\begin{definition}[Spectral simplex]
Let
\[
\mathcal{S}
=
\Delta^{N-1}
=
\left\{
\lambda \in \mathbb{R}^N
\;\middle|\;
\lambda_i \ge 0,\;
\sum_{i=1}^N \lambda_i = 1
\right\}.
\]
We refer to $\mathcal{S}$ as the \emph{spectral simplex}.
\end{definition}

\noindent Each non-zero imaging operator induces a point $\lambda(O)\in\mathcal{S}$.

\subsection{Interior and boundary}
\medskip
\noindent
\begin{proposition}[Rank correspondence]
\label{prop:rank_correspondence}
Let $O$ be an imaging operator with rank $r$.
\begin{itemize}
\item If $r=N$, then $\lambda(O)$ lies in the interior of $\mathcal{S}$.
\item If $r<N$, then $\lambda(O)$ lies on a boundary face of $\mathcal{S}$ of codimension $N-r$.
\end{itemize}
\end{proposition}

\subsection{Spectral equivalence}
\medskip
\begin{definition}[Spectral equivalence]
Two imaging operators $O_1$ and $O_2$ are said to be \emph{spectrally equivalent} if
\[
\lambda(O_1)=\lambda(O_2).
\]
\end{definition}

\medskip
\noindent
The construction above fixes a modelling choice: squared singular values are normalised and interpreted as a probability vector on the simplex. No claim is made that this choice is unique or inevitable. Rather, it provides a minimal operator-invariant state description on which a standard information geometry may be defined.

\section{Information Geometry on the Spectral Simplex}
\label{sec:geometry}
\medskip
\subsection{Fisher--Rao metric}
\medskip
\begin{definition}[Fisher--Rao metric]
On the interior $\mathcal{S}_{>0}$ of the spectral simplex, define the Riemannian metric \cite{AmariNagaoka2000Methods}
\[
g_{ij}(\lambda)=\frac{\delta_{ij}}{4\lambda_i},
\qquad \lambda_i>0.
\]
\end{definition} 

\medskip
\noindent\textbf{Domain convention.}
All geometric statements in this paper are made on the smooth interior manifold $\mathcal{S}_{>0}$, where $\lambda_i>0$ for all $i$.
When boundary faces (i.e.\ $\lambda_i=0$ for some $i$) are discussed, this refers only to limiting behaviour of interior quantities as $\lambda$ approaches the boundary. Boundary points themselves are not elements of the Riemannian manifold, and no metric tensor is defined there. This is the standard Fisher--Rao metric on the simplex \cite{AmariNagaoka2000Methods}.

\medskip
\noindent\textbf{Scaling convention.}
The factor $1/4$ is a fixed normalisation chosen so that the square-root embedding
$\phi(\lambda)=(\sqrt{\lambda_1},\ldots,\sqrt{\lambda_N})$ yields the distance formula in Proposition~\ref{prop:fr_distance} with the prefactor $2$.
All geometric statements below are invariant under constant rescaling of the metric, up to the corresponding rescaling of distances and curvature scale.

\subsection{Distance and embedding}
\medskip
\begin{proposition}[Fisher--Rao distance]
\label{prop:fr_distance}
The geodesic distance induced by $g_{ij}$ is \cite{AmariNagaoka2000Methods}
\[
d_{\mathrm{FR}}(\lambda,\mu)
=
2\arccos\!\left(
\sum_{i=1}^N \sqrt{\lambda_i\mu_i}
\right).
\]
\end{proposition}

\begin{proposition}[Isometric embedding]
\label{prop:isometric_embedding}
The map $\phi:\mathcal{S}_{>0}\to S^{N-1}_{>0}$ defined by $\phi(\lambda)=(\sqrt{\lambda_1},\ldots,\sqrt{\lambda_N})$ is an isometric embedding (up to a constant factor) of $(\mathcal{S}_{>0},g_{\mathrm{FR}})$ into the unit sphere with the round metric \cite{AmariNagaoka2000Methods}.
\end{proposition}

\subsection{Geodesics}
\medskip
\begin{proposition}[Geodesics]
\label{prop:geodesics}
For $\lambda,\mu\in\mathcal{S}_{>0}$, the unique Fisher--Rao geodesic $\gamma:[0,1]\to\mathcal{S}_{>0}$ joining them is
\[
\gamma_i(t)
=
\frac{\big((1-t)\sqrt{\lambda_i}+t\sqrt{\mu_i}\big)^2}
{\sum_j \big((1-t)\sqrt{\lambda_j}+t\sqrt{\mu_j}\big)^2}.
\]
\end{proposition}

\subsection{Curvature}
\medskip
\begin{proposition}[Constant curvature]
\label{prop:constant_curvature}
Via the square-root embedding of Proposition~\ref{prop:isometric_embedding}, $(\mathcal{S}_{>0},g_{\mathrm{FR}})$ is (up to the fixed scaling in the metric definition) isometric to the positive orthant of a round sphere.
In particular, the Fisher--Rao metric on $\mathcal{S}_{>0}$ has constant positive sectional curvature \cite{AmariNagaoka2000Methods}.
\end{proposition}

\section{Information-Theoretic Functionals}
\label{sec:functionals}
\medskip
Let $H(\lambda)=-\sum_i \lambda_i\log\lambda_i$ denote the Shannon entropy.

\begin{proposition}[Entropy bounds]
\label{prop:entropy_bounds}
For all $\lambda\in\mathcal{S}$ \cite{Shannon1948MathTheory,CoverThomas2006Elements},
\[
0\le H(\lambda)\le \log N.
\]
\end{proposition}

\begin{proposition}[Non-identifiability]
\label{prop:non_identifiability}
The mapping $O\mapsto\lambda(O)$ is not injective; singular-vector information is not retained.
\end{proposition}

\section{Composition-Induced Geometry}
\label{sec:composition}
\medskip
Let $A:\mathcal{H}_0\to\mathcal{H}_1$ and $B:\mathcal{H}_1\to\mathcal{H}_2$ be non-zero operators with $BA\neq 0$.

\medskip
\noindent\textbf{Transport model used in this section.}
Throughout this section, statements about ``composition-induced transport'' on $\mathcal{S}$ are to be understood in a restricted sense.
We analyse (i) rank effects that hold for all linear maps (Propositions~\ref{prop:rank_monotonicity}--\ref{prop:boundary_attraction}), and
(ii) an \emph{idealised aligned transport model} in which composition defines a stage-local map on spectral states only under SVD alignment (Definition~\ref{def:svd_alignment} and Proposition~\ref{prop:aligned_transport}). Outside this aligned setting, $\lambda(BA)$ is well-defined but the mapping
$A \mapsto \lambda(BA)$ does not factor through $\lambda(A)$ in general.
Equivalently, there is no stage-local map
$\Phi_B : \mathcal{S} \to \mathcal{S}$,
determined by $B$ alone, such that
$\lambda(BA) = \Phi_B(\lambda(A))$ for all operators $A$
without retaining singular-vector information.
Accordingly, any language in this section referring to
''transport'', ''contraction'', or ``induced maps''
is restricted strictly to the aligned model described above
and should not be read as a claim about generic operator composition.

\begin{proposition}[Rank monotonicity]
\label{prop:rank_monotonicity}
\[
\mathrm{rank}(BA)\le \min\{\mathrm{rank}(A),\mathrm{rank}(B)\}.
\]
\end{proposition}

\begin{proposition}[Boundary attraction]
\label{prop:boundary_attraction}
If $\mathrm{rank}(B)=r_B<N$, then $\lambda(BA)$ lies on a face of $\mathcal{S}$ of codimension at least $N-r_B$.
\end{proposition}

\begin{definition}[SVD alignment]
\label{def:svd_alignment}
Operators $A$ and $B$ are \emph{SVD-aligned} if the right singular vectors of $B$ coincide with the left singular vectors of $A$.
\end{definition}

\begin{proposition}[Aligned spectral transport]
\label{prop:aligned_transport}
Under SVD alignment and full support,
\[
\lambda_i(BA)
=
\frac{\beta_i^2\,\lambda_i(A)}{\sum_j \beta_j^2\,\lambda_j(A)},
\qquad
\beta_i=\sigma_i(B).
\]
\end{proposition}

\begin{proposition}[Isometries]
\label{prop:isometries}
In the aligned case, $\Phi_B$ is an isometry of $(\mathcal{S}_{>0},g_{\mathrm{FR}})$ if and only if $\sigma_1(B)=\cdots=\sigma_N(B)$.
\end{proposition}

\begin{proposition}[Distance distortion under aligned transport]
\label{prop:distance_distortion}
Under the aligned transport model of Proposition~\ref{prop:aligned_transport}, the induced map on $\mathcal{S}_{>0}$ preserves Fisher--Rao distances if and only if $B$ is spectrally uniform (Proposition~\ref{prop:isometries}).
In particular, if $B$ is not spectrally uniform, the aligned transport map is not an isometry and therefore distorts Fisher--Rao distances on $\mathcal{S}_{>0}$.
\end{proposition}

\section{Demonstrative Examples}
\label{sec:demonstrations}
\medskip
This section provides a small number of simple, schematic examples intended solely to illustrate how operators are represented as points and paths in the spectral simplex $\mathcal{S}$ under the geometric construction developed above. 

\medskip
\noindent All examples are analytic and abstract. They are explicitly \emph{demonstrative} and do not constitute general results, validation of the framework, or evidence of typical behaviour in practical systems. No new definitions or claims are introduced in this section.

\subsection{Rank-one operator}
\medskip
Consider an operator $O$ with exactly one non-zero singular value $\sigma_1>0$ and $\sigma_i=0$ for $i>1$. The associated spectral state is
\[
\lambda(O) = (1,0,\ldots,0),
\]
corresponding to a vertex of the spectral simplex $\mathcal{S}$.

\medskip
\noindent
This example illustrates the representation of rank-deficient operators as boundary points of $\mathcal{S}$.

\medskip
\noindent As established earlier, such points lie outside the smooth interior manifold $\mathcal{S}_{>0}$ on which the Fisher--Rao metric is defined.
Geodesics defined by the Fisher--Rao metric remain in $\mathcal{S}_{>0}$ for all $t\in[0,1]$, so boundary points are treated only as limiting endpoints rather than as points within the Riemannian manifold. No interpretation beyond this geometric placement is implied.

\subsection{Spectrally uniform operator}
\medskip
Let $O$ be an operator with singular values $\sigma_1=\cdots=\sigma_N>0$. The corresponding spectral state is
\[
\lambda(O) = \left(\tfrac{1}{N},\ldots,\tfrac{1}{N}\right),
\]
which is the barycentre of $\mathcal{S}$.

\medskip
\noindent
This example demonstrates the central point of the simplex and provides a reference for geometric symmetry. As shown previously, spectrally uniform operators generate isometries under aligned composition; here this property is not re-derived, merely illustrated by the location of the state at the geometric centre.

\subsection{Two-mode operator family}
\medskip
Consider a one-parameter family of operators with two non-zero singular values,
\[
\sigma_1(t)=\cos t,\qquad \sigma_2(t)=\sin t,
\]
with $t\in(0,\tfrac{\pi}{2})$, and all remaining singular values zero. After normalisation, the spectral state is
\[
\lambda(t)=\big(\cos^2 t,\sin^2 t,0,\ldots,0\big).
\]

\noindent As $t$ varies, $\lambda(t)$ traces a curve along a one-dimensional boundary face of $\mathcal{S}$. This curve lies entirely on the boundary and, therefore, does not constitute a Fisher--Rao geodesic in the interior geometry. The example serves only to visualise motion constrained to a lower-dimensional face induced by fixed rank.

\subsection{Interior geodesic between full-rank states}
\medskip
Let $\lambda^{(0)},\lambda^{(1)}\in\mathcal{S}_{>0}$ be two strictly positive spectral states. The Fisher--Rao geodesic connecting them is given explicitly by Proposition~\ref{prop:geodesics}. 

\medskip
\noindent
This demonstration highlights that interior geodesics preserve full support for all $t\in[0,1]$ and, therefore, represent continuous redistributions of spectral weight without rank change. No claim is made that such geodesics correspond to physical or algorithmic deformations of operators; they are shown only as geometric paths in state space.

\subsection{Aligned re-weighting under composition}
\medskip
Consider an operator $A$ with spectral state $\lambda(A)\in\mathcal{S}_{>0}$ and an aligned operator $B$ with singular values $\{\beta_i\}_{i=1}^N$, all strictly positive. Under the aligned composition model, the resulting spectral state is
\[
\lambda_i(BA)=\frac{\beta_i^2\,\lambda_i(A)}{\sum_j \beta_j^2\,\lambda_j(A)}.
\]

\noindent This example illustrates composition as a smooth, non-linear re-weighting of spectral coordinates followed by renormalisation. When the $\beta_i$ are not all equal, the resulting path in $\mathcal{S}_{>0}$ is generally not a geodesic and distorts distances relative to the Fisher--Rao metric. The example is included solely to visualise how composition may move spectral states within the simplex under a specific, idealised alignment assumption.

\subsection{Boundary attraction by rank-deficient stages}
\medskip
Let $B$ be an operator of rank $r<N$, and let $A$ be any operator such that $BA\neq 0$. As shown in Section~\ref{sec:composition}, the spectral state $\lambda(BA)$ lies on a face of $\mathcal{S}$ of codimension at least $N-r$.

\medskip
\noindent
This demonstration illustrates the geometric effect of rank reduction as a projection onto the boundary of the simplex. No claim is made regarding the reversibility, typicality, or physical interpretation of this effect; it is shown only as a schematic consequence of the rank constraint.

\subsection{Summary of demonstrations}
\medskip
The examples above are intended to provide geometric intuition for the abstract constructions developed in this paper. They illustrate how simple operator families appear as points, curves, and boundary limits in the spectral simplex, and how composition may act as transport within this space under restrictive assumptions. No general behaviour, performance claims, or empirical validation should be inferred from these demonstrations.

\section{What This Geometry Does and Does Not Say}
\label{sec:limits_interpretation}
\medskip
This section delineates the precise scope of the information geometry introduced in this paper. Its purpose is to prevent over-interpretation of the formal results, and to clarify: which quantities are meaningful within the construction, which properties are invariant, which questions are explicitly excluded. No new mathematics is introduced.

\subsection{What the geometry does say}

\medskip
\noindent
The geometry defined on the spectral simplex $\mathcal{S}$ makes the following statements precise:

\paragraph{Invariant quantities.}
The construction yields invariants of imaging operators under unitary pre- and post-composition and global rescaling. In particular:
\begin{itemize}
\item the spectral state $\lambda(O)$,
\item Fisher--Rao distances between spectral states,
\item geodesic structure within $\mathcal{S}_{>0}$,
\item curvature properties of the interior geometry,
\end{itemize}
depend only on the normalised singular spectrum of the operator. These quantities, therefore, characterise equivalence classes of operators that are informationally indistinguishable at the level of spectral energy distribution.

\paragraph{Meaningful comparisons.}
Distances in the Fisher--Rao geometry quantify differences in relative spectral allocation and are invariant to basis choice and global scaling.

\paragraph{Rank structure and boundaries.}
The geometry distinguishes sharply between full-rank and rank-deficient operators. Full-rank operators correspond to interior points of $\mathcal{S}$ and admit smooth geodesic connections. Rank-deficient operators lie on boundary faces, which are metrically singular. This distinction is intrinsic to the geometry and does not depend on reconstruction, discretisation refinement, or algorithmic choices.

\paragraph{Effects of composition.}
For linear stages represented within a fixed discretisation, composition has two distinct kinds of consequences in this paper.
First, rank monotonicity and boundary attraction (Propositions~\ref{prop:rank_monotonicity}--\ref{prop:boundary_attraction}) hold for arbitrary linear maps.
Second, under the explicitly stated SVD-aligned transport model (Definition~\ref{def:svd_alignment} and Proposition~\ref{prop:aligned_transport}), composition induces a non-linear re-weighting on $\mathcal{S}_{>0}$ that is an isometry if, and only if, the composing stage is spectrally uniform (Proposition~\ref{prop:isometries}).
No claim of a canonical transport map, contraction, or curvature evolution is made for general non-aligned composition.

\subsection{What the geometry does not say}
\medskip
The construction deliberately excludes a number of questions that might otherwise be inferred.

\paragraph{No optimisation principle.}
The geometry does not define optimal operators, optimal pipelines, or preferred directions of evolution. Geodesics describe shortest paths in the chosen metric, not physically realised or desirable transformations. No claim is made that imaging systems evolve along geodesics or should be designed to do so.

\paragraph{No performance or quality metric.}
Distances in the spectral simplex do not measure image quality, reconstruction fidelity, or task performance. Proximity in this geometry reflects similarity of spectral distributions, not superiority of one operator over another.

\paragraph{No stochastic interpretation.}
Although the Fisher--Rao metric originates in statistical geometry, the present construction does not model noise, uncertainty, or probability distributions over data. Any contraction or monotonicity properties associated with stochastic maps are not asserted for general operator composition.

\paragraph{No statement about singular vectors.}
The geometry is intentionally blind to singular-vector structure. It does not encode spatial localisation, mode alignment, commutativity, or directional information. Operators that differ substantially in their action on object space may coincide as points in $\mathcal{S}$ if their normalised singular spectra agree.

\paragraph{No physical modality or mechanism.}
The geometry is independent of imaging modality, physical interaction, and measurement mechanism. It arises entirely from the abstract representation of operators through their singular spectra and should not be interpreted as a model of physical space, propagation, or dynamics.

\subsection{Preventing over-interpretation}
\medskip 
The geometry developed here is a structural constraint, not a descriptive model of physical processes or algorithms. Its role is to fix what it means for operators to be informationally similar, deformable, or singular at the spectral level. Any attempt to infer optimisation strategies, physical dynamics, or algorithmic guarantees directly from this geometry goes beyond the claims of the present paper. Questions concerning noise models, reconstruction algorithms, physical design, spatio-temporal dynamics, or emergent behaviour require additional structure and are deferred to subsequent work. The present results should, therefore, be read as establishing a reference geometry against which such extensions may be formulated, rather than as providing answers to those questions themselves.

\section{Discussion}
\label{sec:discussion}
\medskip 
The purpose of this paper has been to identify and fix a minimal geometric structure induced by imaging operators when viewed through their normalised singular spectra. The discussion here synthesises the consequences of that construction without restating definitions or technical results, and clarifies why the deliberate minimality of the framework is essential.

\medskip
\noindent
Once imaging operators are reduced to their spectral equivalence classes, an information geometry arises unavoidably. The spectral simplex provides the natural state space, and the Fisher--Rao metric supplies a canonical notion of distance, curvature, and geodesic structure on that space. This geometry is invariant under unitary transformations and global rescaling, and, therefore, depends only on how an operator distributes energy across its singular modes. In this sense, the geometry captures what is common to all representations of an operator that are informationally indistinguishable at the spectral level.

\medskip
\noindent
Minimality is not a limitation; it's a structural requirement. By restricting attention to normalised singular spectra, the construction avoids embedding assumptions about noise models, reconstruction algorithms, physical modality, or optimisation criteria. As a result, the geometry isolates those aspects of imaging operators that are unavoidable consequences of linear measurement itself. Any richer description (incorporating singular vectors, stochastic effects, non-linearities, or task-specific objectives) must be built on top of this structure rather than modifying it. The geometry, therefore, functions as a fixed background against which additional structure can be meaningfully introduced and compared.

\medskip
\noindent
A key consequence of this minimal construction is the sharp distinction it enforces between interior and boundary behaviour. Full-rank operators occupy the interior of the spectral simplex and admit smooth geometric relations, while rank-deficient operators lie on metrically singular boundary faces. This boundary structure is intrinsic to the geometry and reflects algebraic loss of rank under fixed discretisation, corresponding to vanishing spectral coordinates. The appearance of boundary attraction under composition is not an artefact of modelling choices, but a geometric manifestation of rank reduction and mode elimination. The term ``irreversible'' is used here only in this algebraic sense: no statement is made about numerical effective rank, noise robustness, regularisation, or practical reconstructability in applied imaging systems.

\medskip
\noindent
The analysis of composition further clarifies what can, and cannot, be expected from a purely spectral geometry. Composition generally induces metric distortion and curvature effects, and only in the spectrally uniform case does it act as an isometry. No general contraction, monotonicity, or geodesic flow property holds without additional assumptions. This absence of global simplification is itself a structural result: it shows that information loss and redistribution under composition cannot, in general, be summarised by a single scalar or monotone quantity at the spectral level alone.

\medskip
\noindent
By fixing a well-defined notion of distance and curvature between spectral states, the present work establishes a common reference frame for comparing imaging operators abstractly. Importantly, this comparison is neither task-based nor performance-based. Proximity in the spectral geometry does not imply similarity in reconstruction quality or physical behaviour; it implies similarity in how information is allocated across modes. This distinction prevents over-interpretation and ensures that the geometry is used as a structural constraint rather than a proxy for optimisation or evaluation.

\medskip
\noindent
In summary, this paper does not introduce new methods for designing imaging systems or analysing data. It fixes a geometric structure that any such analysis must respect once operators are compared at the level of their singular spectra. The structure is minimal, invariant, and mathematically rigid. Its value is to provide a common reference geometry for operator comparison: by removing contingent modelling choices, it makes explicit which aspects of information allocation are imposed by operator structure alone.

\section{Conclusion}
\label{sec:conclusion}
\medskip 
This paper has established a minimal information-geometric structure induced by imaging operators when represented through their normalised singular spectra. By identifying spectral equivalence classes of operators with points on the probability simplex and equipping this space with the Fisher--Rao metric, we have fixed a canonical geometry that is invariant under unitary transformations and global rescaling. Within this framework, notions of distance, geodesic connection, curvature, and boundary behaviour are well-defined and arise without additional modelling assumptions.

\medskip
\noindent
By fixing this geometric background, the paper closes a degree of freedom that is often left implicit in discussions of imaging operators and information flow. From this point onward, any comparison, deformation, or analysis of operators at the spectral level necessarily refers to the same underlying geometry. Subsequent papers in this programme will build on this fixed structure by introducing additional layers of description where appropriate, but the geometric foundation established here is intended to remain unchanged.

\medskip
\noindent
The contribution of this work is, therefore, foundational rather than expansive: it delineates a precise, invariant setting in which operator-level information can be compared and constrained, while deliberately refraining from claims that extend beyond that setting.

\ack{None.}

\funding{No specific funding was received for this work.}

\roles{The author conceived the study, developed the theoretical framework, designed and performed the analyses, and wrote the manuscript.}

\data{No new data were created or analysed in this theoretical study.}

\suppdata{No supplementary material is provided.}

\bibliographystyle{iopart-num} 

\bibliography{References/references}    

\end{document}